\documentclass[reprint,prd,twocolumn,preprintnumbers,superscriptaddress,dblfloatfix]{revtex4-1}
\usepackage{graphicx}   
\usepackage{dcolumn}    
\usepackage{bm}         
\usepackage{amssymb}    
\usepackage{setspace}
\usepackage{amsmath, amssymb, setspace}
\usepackage{array}
\usepackage{booktabs}
\usepackage{mathrsfs}
\usepackage{indentfirst}
\usepackage{slashed}
\usepackage{float}
\usepackage{lmodern}
\usepackage{hyperref}
\usepackage{xcolor}
\usepackage{multirow}
\usepackage{epstopdf}
\usepackage{ulem}
\usepackage{tabularx}
\usepackage[justification=raggedright]{caption}
\usepackage[flushleft]{threeparttable}
\usepackage{hyperref}
\usepackage{soul}
\usepackage[utf8]{inputenc}

\begin{document}

\preprint{IPMU19-0115}

\title{Fundamental Forces and Scalar Field Dynamics in the Early Universe}

\author{Alexander Kusenko}
\email[]{kusenko@ucla.edu}
\affiliation{Department of Physics and Astronomy, University of California, Los Angeles\\
Los Angeles, California, 90095-1547, USA}
\affiliation{Kavli Institute for the Physics and Mathematics of the Universe (WPI), UTIAS\\
The University of Tokyo, Kashiwa, Chiba 277-8583, Japan}

\author{Volodymyr Takhistov}
\email[]{vtakhist@physics.ucla.edu}
\affiliation{Department of Physics and Astronomy, University of California, Los Angeles\\
Los Angeles, California, 90095-1547, USA}

\author{Masaki Yamada}
\email[]{masaki.yamada@tufts.edu}
\affiliation{Institute of Cosmology, Department of Physics and Astronomy, Tufts University, \\ 
574 Boston Avenue, Medford, Massachusetts, 02155, USA}

\author{Masahito Yamazaki}
\email[]{masahito.yamazaki@ipmu.jp}
\affiliation{Kavli Institute for the Physics and Mathematics of the Universe (WPI), UTIAS\\
The University of Tokyo, Kashiwa, Chiba 277-8583, Japan}

\date{\today} 

\begin{abstract} 
Scalar weak gravity conjectures (SWGCs) attempt to pinpoint the ranges of couplings consistent with a fundamental theory of all interactions.  We identify a generic dynamical consequence of these conjectures for cosmology and show that SWGCs imply a particular behavior of the scalar fields in the early universe.  A scalar field that develops a large expectation value during inflation must relax to the minimum of effective potential at a later time. SWGCs imply that a homogeneous distribution of the field is unstable with respect to fragmentation into localized lumps, which could potentially lead to significant consequences for cosmology.
\end{abstract}

\maketitle

While string theory has potential to explain all forces of Nature, the large number of possible vacua has so far stymied the efforts to connect its predictions to the Standard Model (SM). However, it appears that a significant portion of the observable parameter space of low-energy effective field theories (EFTs) belongs to the ``swampland"~\cite{Vafa:2005ui,Ooguri:2006in} region that is forbidden in more fundamental quantum-gravity theory (see e.g.~Ref.~\cite{Brennan:2017rbf,Palti:2019pca,Yamazaki:2019ahj} for summary).  In this {\it Letter} we will describe a generic pattern of dynamics in the early universe that is implied by some of the proposed conjectures.

It has been suggested that some relations between coupling constants, such as those describing the relative strengths of gravity and other interactions, must be satisfied in the low-energy EFT. 
A quantum theory of gravity appears to require that gravity must be ``weaker" than any gauge interaction~\cite{ArkaniHamed:2006dz}. While the original ``weak gravity conjecture'' (WGC)~\cite{ArkaniHamed:2006dz} referred only to gravity and gauge interactions, it has been proposed that interactions mediated by scalar fields must also be stronger than gravity in appropriate units~\cite{Palti:2017elp,Gonzalo:2019gjp}. 
There are several versions of this conjecture, which we collectively refer to as ``scalar weak gravity conjectures'' (SWGCs). Further studies are necessary to identify which particular formulation best captures the self-consistency conditions arising from a fundamental theory. 

We will show that various proposed  versions of SWGC appear to have a common set of implications for the early universe cosmology, and we will identify some common patterns of behavior for the scalar fields.
Scalar fields are known to exist in nature, as confirmed by the discovery of the Higgs boson. Many additional scalar fields are predicted by string theory.  We will see that, if SWGCs hold, then in an expanding universe a broad range of scalar fields evolve into inhomogeneous configurations consisting of isolated solitonic lumps of scalar condensate.  While such configurations may be short-lived, their formation can have profound implications. In particular, primordial black holes (PBHs) or stable field lumps can form and can presently exist as a form of dark matter (DM). We will show that the conditions for the early-universe inhomogeneities of scalar fields are parametrically similar, if not identical, to the constraints arising from SWGCs. The basic reason for this connection has to do with attractive forces: SWGCs imply that scalar-mediated forces are attractive and stronger than gravity. In turn, the attractive forces lead to instabilities that cause an initially homogeneous solution to fragment into lumps. 

There is a strong support for the idea that the early universe underwent a stage of inflation, during which the space-time was approximately de Sitter, and the energy density was dominated by the inflaton field slowly rolling toward the minimum of its effective potential from some initial condition, often assumed to be random/chaotic.  While the inflaton plays a singularly important role, the rest of the scalar fields get engaged in a non-trivial dynamical behaviour. Namely, a light spectator scalar field $\phi$ acquires a non-zero expectation value $\langle \phi \rangle $ during inflation, such that the energy density in each scalar degree of freedom is $U(\langle \phi \rangle)\sim H_I^4$, where $H_I$ is the Hubble parameter during inflation~\cite{Bunch:1978yq,Linde:1982uu,Lee:1987qc,Starobinsky:1994bd,Dine:2003ax,Cotner:2019ykd}.  

When inflation is over, the scalar field $\phi$ must evolve toward the minimum of its effective potential. While the initial condition set by inflation is $\phi \approx {\rm const} $ on superhorizon scales, in general, the field does not remain homogeneous due to an instability associated with attractive self-interaction forces~\cite{Kusenko:1997si,Dine:2003ax}.  In contrast, repulsive forces assist in stabilizing the homogeneous solution, as illustrated in the (unphysical) example of repulsive gravity~\cite{Guth:2014hsa}.

Gravitational force is attractive, and it causes a well-known  instability that renders a homogeneous distribution of mass unstable to  fragmentation.  However, for a  {\it spectator} field, which does not significantly contribute to the energy density, the gravitational instability is ineffective.  Even if a scalar field comes to dominate the energy density, the gravitational instability has a limited effect.
On superhorizon length scales, the instability cannot grow due to causality. On smaller scales, the fate of the instability depends on the presence of any repulsive self-interactions. In the absence of any repulsion, for example in the case of a universe filled with pressureless gas, density perturbations grow on subhorizon scales. In the case of an atomic hydrogen gas, the pressure is non-zero, but its effects are communicated with the speed of sound, which is much smaller than the speed of light. The effects of pressure are important on the scales smaller than the sound horizon known as the Jeans scale.  On the scales greater than the Jeans length, fragmentation and collapse ensue.  In the case of radiation, the pressure is non-zero (and is of the same order or magnitude as the energy density), and the speed of sound is close to the speed of light.  In this case, the pressure prevents the instability from developing on subhorizon scales, while causality prevents it from developing on superhorizon scale.  The latter has to do with the fact that gravity controls both the rate of gravitational instability and the expansion rate, with which the instability competes. 

In the case of a scalar field, there is a possibility of an attractive force unrelated to gravity (and, therefore, unrelated to the expansion rate). The self-interaction couples to the occupation density of scalar quanta and not the energy density. Hence, the instability can be fast and effective even for a spectator field~\cite{Kusenko:1997si}.  There is also a possibility of repulsive self-interaction.  If attractive self-interaction dominates over repulsion, and if it is strong enough to counter the effects of expansion of the universe, then a homogeneous field configuration is unstable with respect to fragmentation.  It is intriguing that {\it SWGCs impose the same conditions on the scalar field, namely, the dominance of attractive self-interaction over repulsion and over gravity}. 

When attractive self-interactions are present, a time-dependent spatially homogeneous solution is often found to be unstable, and the field fragments into solitonic lumps~\cite{Kusenko:1997si}, known as Q-balls~\cite{Coleman:1985ki}. The case of real scalar field fragmenting into oscillons is similar.  Fragmentation can be analyzed by finding the unstable modes~\cite{Kusenko:1997si} or by identifying the regime of negative pressure~\cite{Turner:1983he,McDonald:1993ky}. For example, for an effective potential $V_{\rm eff} \propto |\phi|^n$ with $n < 2$, the self-interaction creates  negative pressure and  leads to fragmentation  into field lumps~\cite{Enqvist:1997si}.  The condition $n < 2$ corresponds to the potential energy rising slower than that for non-interacting scalar field ($n=2$), which means the self-interactions are attractive. 

Nearly flat potentials arise along the flat directions in supersymmetric extensions of the standard model, and such potentials admit Q-balls~\cite{Coleman:1985ki}, which are the final state of fragmentation~\cite{Kusenko:1997si}.  
An illustrative example is a polynomial potential in which the cubic term represents the attractive interaction, while the quartic term is repulsive:
\begin{equation} \label{eq:mainpot}
V(\phi) = \dfrac{1}{2} m^2 \phi^2 - \dfrac{A}{3}\phi^3 + \dfrac{\lambda}{4}\phi^4~.
\end{equation}
Potentials of this kind can appear in theories with supersymmetry.  For example, in the Minimal Supersymmetric Standard Model (MSSM), there are tri-linear terms of the form $\tilde{H}\tilde{Q}\tilde{q}$ where $\tilde{H}$ is the Higgs field, $\tilde{Q}$ is a squark SU(2) doublet and $\tilde{q}$ is the squark singlet.  In addition, there are quartic terms of the form $\tilde{H}^\dag \tilde{H} \tilde{Q}^\dag \tilde{Q}$, $|\tilde{Q}^\dag \tilde{Q}|^2$, etc.  The potential respects the global $\mathrm{U}(1)$ symmetry corresponding to the baryon number conservation.  Parameterized in terms of some linear combination of $\tilde{H}$, $\tilde{Q}$, $\tilde{q}$ denoted as $\phi$, the MSSM scalar potential takes the form of Eq.~(\ref{eq:mainpot}) in the $\phi$ direction. Below we discuss application of SWGCs to potential of Eq.~\eqref{eq:mainpot}.

SWGCs can be formulated in terms of the derivatives $V'', V''', V''''$ of the potential $V$, which are controlled by the coefficients 
$m, A, \lambda$, respectively. 
We have chosen $A\ge 0$ without loss of generality, since the sign can be changed by the transformation $\phi \rightarrow -\phi$. 
Since there is always a direction in which the cubic term is negative, the coupling $A$ contributes to bringing the potential below the non-interacting case $V\propto \phi^2$, and, therefore, the cubic term represents an attractive interaction. 
Similarly, $\lambda>0$ represents a repulsive force generated by the scalar field, since the term $\phi^4$ grows faster than $\phi^2$.

Q-ball solutions exist if 
the homogeneous distribution is not the minimum of energy at fixed global charge, or, equivalently, when there exists some field value $\phi_0 \neq 0$ such that~\cite{Coleman:1985ki}  
\begin{equation}
 \frac{V(\phi_0)}{\phi_0^2/2} < m^2\equiv V''(0)~~\Longrightarrow~~~ A > \dfrac{3 \lambda \phi_0}{4}~.
 \label{eq:condition}
\end{equation}
For any positive $A$, there is a value $\phi_0$ that satisfies this inequality, which implies that Q-balls exist whenever $A > 0$. 
In the following analysis, we assume $A^2 < 9 \lambda m^2 /2$ so that $\phi = 0$ is the global minimum. In this case, Q-balls exist for any charge $Q$. In the false vacuum there is a maximal charge beyond which the Q-ball expands and causes a phase transition~\cite{Kusenko:1997hj}. The condition of Eq.~\eqref{eq:condition} can be generalized to the case of multiple fields~\cite{Kusenko:1997si}.

Instability of a homogeneous solution can be studied by means of Floquet analysis~\cite{Kusenko:1997si}.
The stability of the homogeneous solution $\phi = \Phi(x,t)e^{i \Omega(x,t)}$ can be analyzed by adding a  small perturbation $\delta \Phi$, $\delta \Omega \propto e^{S(t) - \vec{k} \vec{x}}$ and searching  for growing modes with ${\rm Re}(\alpha) \equiv {\rm Re}(dS/dt)>0$. The dispersion relations~\cite{Kusenko:1997si,Doddato:2011fz} imply that the fastest growing mode for the potential~(\eqref{eq:mainpot})  
has the Floquet exponent  
\begin{align}
\alpha_{\rm max}
= \dfrac{\phi(A - 2 \lambda \phi)}{4 \sqrt{m^2 - A \phi + \lambda \phi^2}}~.
\end{align}
Assuming the energy density in the $\phi$ field is smaller but not significantly less than the critical density, $V(\phi) = f \rho_c$, where $f<1$, the Hubble parameter is  $H \sim m \phi/ \sqrt{6f} M_{\rm Pl}$. 
The instability has time to develop and become nonlinear if $\alpha_{\rm max}> H$. 
Comparing $H$ with $\alpha_{\rm max}$, 
one obtains the condition for the instability to overcome the expansion: 
\begin{equation} \label{eq:instab}
\dfrac{\alpha_{\rm max}}{H} 
\sim \dfrac{A M_{\rm Pl}}{m^2} \gtrsim 1~,
\end{equation}
where we have assumed that $f$ is not much smaller than one ($f\lesssim  1$) and used Eq.~\eqref{eq:condition} and the fact that $\lambda \phi^2,A \phi\lesssim m^2$ (we note that this is correct to an $\mathcal{O}(1)$ uncertainty, even if $\lambda \phi^2$ and $A \phi$ are as large as $\mathcal{O}(m^2)$).
As shown by both analytic calculations and numerical simulations~\cite{Kusenko:1997si, Kasuya:2000wx, Multamaki:2002hv}, the unstable growing modes eventually enter a non-linear regime and fragment into Q-balls. 

The case of real scalar field is analogous. Let us consider a real scalar field $\phi$ with potential \eqref{eq:mainpot} as before.
In the non-relativistic limit, the Lagrangian can be written as~\cite{Mukaida:2016hwd}
\begin{equation}
    \begin{cases}
L_{\rm eff} = \dfrac{1}{4}\Big[\Phi^{\dagger}(2 i m \partial_t - \partial_t^2 +    \nabla^2) \Phi - V_{\rm eff} - i \Gamma \Big] ~,\\  
V_{\rm eff} \simeq \Big(-\dfrac{5 A^2}{12 m^2} + \dfrac{3 \lambda}{8}\Big) |\Phi|^4 + \dfrac{\lambda^2}{128 m^2}|\Phi|^6 ~,
    \end{cases}
\end{equation}
where $\phi = (\Phi e^{-i m t} + {\rm h.c.})/2$ with complex $\Phi$.
This theory has an approximate $\mathrm{U}(1)$-symmetry rotating the phase of $\Phi$,
and $\Gamma$ represents the $\mathrm{U}(1)$-breaking imaginary terms, which we neglect. Then, we can identify a Q-ball solution for $\Phi$ when the coefficient of $|\Phi|^4$ is negative:
\begin{equation} \label{eq:condosc}
 \lambda < \frac{10 A^2}{9 m^2}~.
\end{equation}
We note that the mass term is hidden within the kinetic terms.~The oscillon solution $\phi(x)$ is nothing else but a real component projection of the Q-ball solution $\Phi(x,t)e^{-i m t}$. 
The same condition can be derived by using the $\varepsilon$-expansion method as done in Ref.~\cite{Fodor:2008es}, where their condition $\lambda >0$ is identical to our Eq.~(\ref{eq:condosc}). 
Calculating the Floquet exponent for small $\phi$ using the method of  Ref.~\cite{Amin:2010xe}, one can verify that Eq.~(\ref{eq:instab}) is the condition for the instability to grow faster than the expansion of the Universe.  Thus the interplay of 
scalar-mediated forces and gravity determines the cosmological history of the scalar condensate. 

The WGC~\cite{ArkaniHamed:2006dz} states that in a theory compatible with quantum gravity ultraviolet (UV) completion there is a particle $m$ with $\mathrm{U}(1)$ charge $q$ satisfying 
$\sqrt{2} q e \geq m/M_{\rm Pl}$,
where $e$ is the $\mathrm{U}(1)$ gauge coupling.
This statement is supported by substantial circumstantial evidence and arguments related to black holes~\cite{ArkaniHamed:2006dz} (see also, e.g.,~Ref.~\cite{Cheung:2014ega,Harlow:2015lma,Cottrell:2016bty,Hod:2017uqc,Fisher:2017dbc,Crisford:2017gsb,Cheung:2018cwt,Hamada:2018dde,Urbano:2018kax,Bellazzini:2019xts}). The WGC implies that force mediated by a gauge boson (spin 1) is stronger than that  mediated by the graviton (spin 2), namely 
\begin{equation}
    F_{\rm gauge} ~\Big[= \dfrac{(qe)^2}{4\pi r^2}\Big] \geq F_{\rm gravity} ~\Big[= \dfrac{m^2}{8\pi M_{\rm Pl}^2 r^2}\Big]~.
\end{equation}
Due to association with gauge fields, this can be distinguished as the ``gauge WGC'' (GWGC).
It has been also argued that a version of WGC can be applied to axions. 

SWGC restricts scalar-mediated (spin 0) interactions~\cite{Palti:2017elp,Lust:2017wrl,Lee:2018spm,Heidenreich:2019zkl}. Some versions of the conjecture also involve gauge forces~\cite{Palti:2017elp,Heidenreich:2019zkl} and require that there exists a state such that the repulsive gauge force overcomes the combined attractive forces due to scalar and gravity interactions. We will not discuss
such variants of the conjecture, since a condensate with a non-zero gauge charge is energetically difficult to form, and gauge  Q-ball size is limited due to Coulomb repulsion~\cite{Lee:1988ag}.

In an earlier formulation of SWGC~\cite{Palti:2017elp} there is a particle $\phi$ of mass $m$ coupled to a light scalar $\eta$ with a field-dependent mass $m^2(\eta)|\phi|^2$, resulting in a tri-linear coupling $\partial_{\eta} m$. The SWGC then requires
$2 (\partial_{\eta} m)^2 \geq (m/M_{\rm Pl})^2$, 
which physically means that scalar interactions are stronger than gravity. 
For $m^2 = V''$,  SWGC implies 
\begin{equation}\label{eq:swgc_2}
\frac{1}{2}(V''')^2 \geq  \dfrac{(V'')^2}{M_{\rm Pl}^2} ~.
\end{equation}

The SWGC is on a weaker footing than WGC and is not supported by arguments related to black holes and local symmetries.  In particular, the SWGC as stated in Eq.~\eqref{eq:swgc_2} has been found~\cite{Shirai:2019tgr} to be in phenomenological tension with the swampland de Sitter conjecture~\cite{Obied:2018sgi} as well as its refinements~\cite{Dvali:2018fqu,Andriot:2018wzk,Garg:2018reu,Murayama:2018lie,Ooguri:2018wrx,Garg:2018zdg,Andriot:2018mav}. Some potential counter-examples involve axion-like particles and fifth-force constraints~\cite{Bertotti:2003rm,Will:2005va}.

A different version of SWGC, the strong SWGC (SSWGC)~\cite{Gonzalo:2019gjp}, requires that
\begin{equation} \label{eq:sswgc}
2 (V''')^2 - V''V'''' \geq \dfrac{(V'')^2}{M_{\rm Pl}^2}~.
\end{equation}
The inclusion of quartic contact contribution in Eq.~\eqref{eq:sswgc} makes SSWGC  an ``ultraviolet/infrared (UV/IR) mixing'' statement~\cite{Gonzalo:2019gjp}, and one obtains a tower of states when the equality is saturated. SSWGC is consistent with the SM Higgs, it allows axion-like particles, and it could also avoid the fifth-force searches~\cite{Shirai:2019tgr}.

\begin{figure}[t!]
\includegraphics[angle=0,width=.45\textwidth]{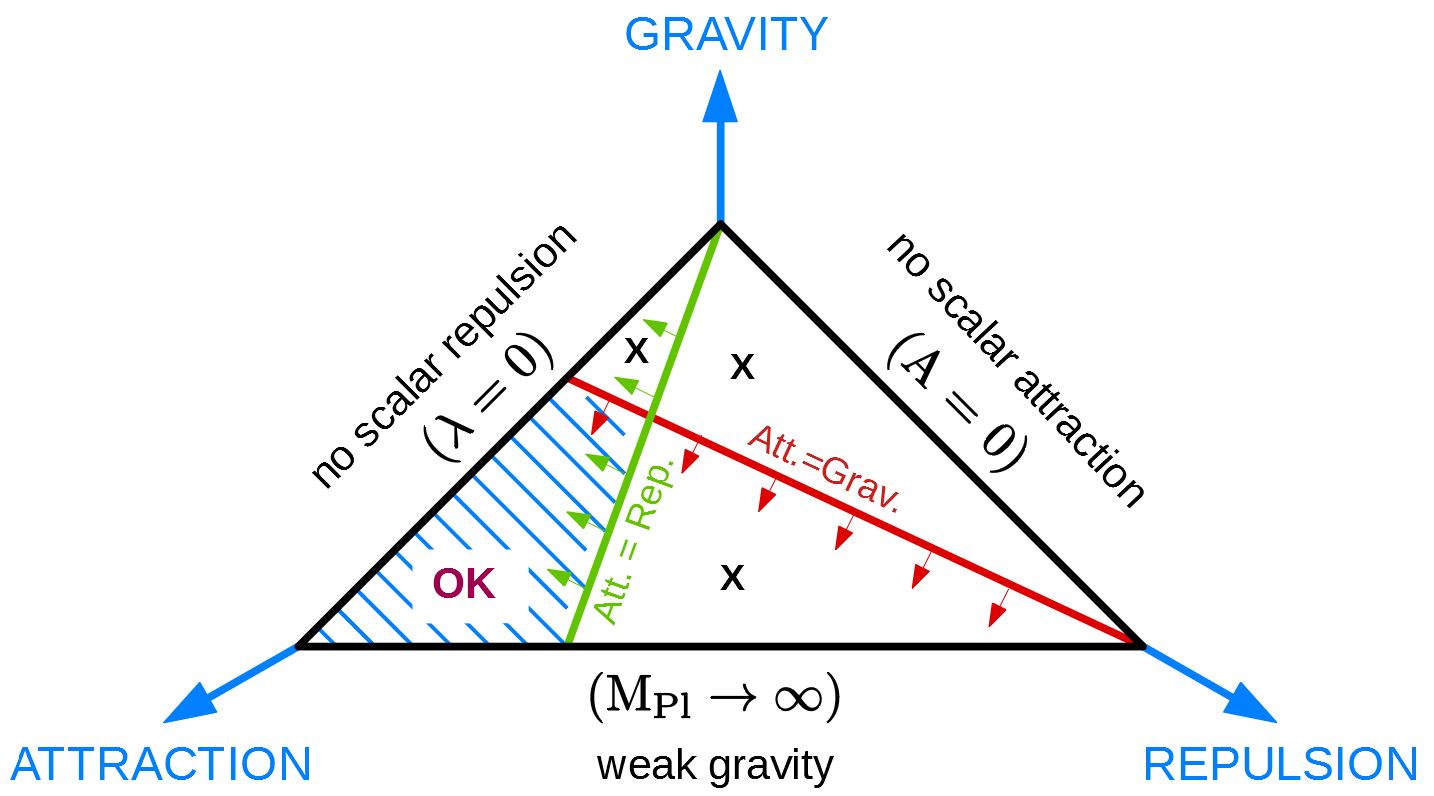}
	\caption{Schematic summary of 
	forces affecting the scalar field behavior in the early universe. Only the region labeled ``OK'' is consistent with GSWGC, and scalar field fragmentation is allowed in the same region.  Lines ``Att.=Grav.'' and ``Att.=Rep.'' correspond to parameters for which the attractive interactions equal gravitational and repulsive interactions, respectively.}
	\label{fig:triangle} 
\end{figure}
\twocolumngrid
At present, evidence supporting the SSWGC is sparse.
In particular, we note that the constraint implied by SSWGC 
($2 (V''')^2 - V''V'''' \geq 0$) remains even in the decoupling limit of gravity $M_{\rm Pl}\to \infty$, in contrast with the WGC, SWGC~\eqref{eq:swgc_2} and other swampland conjectures.

The SSWGC~\eqref{eq:sswgc} has a clear physical interpretation. The first term represents the attractive force mediated by the scalar field  $F_{\rm scalar}^{\rm attraction} \sim (V''')^2$
, while the second term is repulsive force 
$F_{\rm scalar}^{\rm repulsion}\sim V''V''''$.
Hence, the inequality~\eqref{eq:sswgc}
is simply the statement that the 
net attractive force mediated by the scalar field is greater than the force mediated by gravity:
\begin{equation} \label{eq:swg_plain}
F_{\rm scalar}^{\rm net}=
F_{\rm scalar}^{\rm attraction}- F_{\rm scalar}^{\rm repulsion} \gtrsim F_{\rm gravity}~.
\end{equation}
In this language, the inequality of Eq.~\eqref{eq:swgc_2} represents the special case of Eq.~\eqref{eq:swg_plain} when the 
repulsive contributions can be neglected.

The above expressions are just various incarnations of the statement 
that the gravity is the weakest force, and that the scalar attraction is more important.  This has profound implications for the scalar dynamics in expanding universe. 
We emphasize that our analysis does not require that 
conjecture of Eq.~\eqref{eq:sswgc} is a universal statement applicable to 
the whole landscape of the string vacua and our considerations are of value
even if the conjecture holds only in a 
particular corner of the quantum gravity landscape. We also note that our study does not rely on the de Sitter swampland conjecture of Ref.~\cite{Obied:2018sgi} or its refinements~\cite{Dvali:2018fqu,Andriot:2018wzk,Garg:2018reu,Murayama:2018lie,Ooguri:2018wrx,Garg:2018zdg,Andriot:2018mav}.
Moreover, our arguments are not sensitive to the 
specific numerical coefficients in  Eq.~\eqref{eq:sswgc} and will qualitatively hold even if Eq.~\eqref{eq:sswgc}
 is changed to
\begin{equation} \label{eq:gswgc}
2 c (V''')^2 -  c'V''V'''' \geq \dfrac{(V'')^2}{M_{\rm Pl}^2}~,
\end{equation}
where $c$ and $c'$ are some $\mathrm{O}(1)$ positive constants (this inequality reduces to 
SSWGC for $c=c'=1$).  
For definiteness, we use the more general version of the conjecture 
as stated in Eq.~\eqref{eq:gswgc}, which 
we call the generalized scalar weak gravity conjecture (GSWGC).  

To analyze Q-ball and oscillon formation in the context of GSWGC, we apply the conjecture to the potential of Eq.~\eqref{eq:mainpot}. This results in
\begin{align}
8c A^2 - 6c' \lambda m^2 -& 12c A \lambda \phi + 54 c' \lambda^2 \phi^2 \notag\\ \geq& \\ \dfrac{1}{M_{\rm Pl}^2} (m^2 -& 2 A \phi + 3 \lambda \phi^2)^2~, \notag    
\end{align}
which in the regime of $A \phi, \lambda \phi^2 \lesssim m^2$ reduces to
\begin{equation} \label{eq:three_term}
8c A^2 - 6c' \lambda m^2  \gtrsim \dfrac{1}{M_{\rm Pl}^2} (m^2)^2~.    
\end{equation}
This inequality
represents the 
competition between three coefficients $m, A, \lambda$.
It is instructive to consider the limiting cases of Eq.~\eqref{eq:three_term} when two of the terms dominate over the third. This corresponds to the edges of the triangle on Fig.~\ref{fig:triangle} that graphically represents the interplay between coefficients $m$, $A$ and $\lambda$. 

The $(A = 0)$ edge of the triangle represents vanishing attractive scalar interactions, which is not allowed by the condition~of Eq.~\eqref{eq:three_term}, which requires $A>0$. Thus the condition~of Eq.~\eqref{eq:condition} is automatically satisfied, and Q-balls always exist in the theory. In the case $(\lambda \rightarrow 0)$, which represents lack of repulsive scalar interactions,  Eq.~\eqref{eq:three_term} reduces to
\begin{equation}
    A  \gtrsim \dfrac{m^2}{\sqrt{8 c} M_{\rm Pl}}~.
\end{equation}
In view of Eq.~\eqref{eq:instab}, this implies that not only Q-balls exist, but that instabilities can  overcome expansion, and Q-balls (as well as oscillons) can form.
Finally, in the weak gravity limit $(M_{\rm Pl} \rightarrow \infty)$, the condition is 
\begin{equation} \label{eq:sswgcosc}
   \dfrac{4 cA^2}{3 m^2} \gtrsim c' \lambda~.
\end{equation}
From Eq.~(\ref{eq:condosc}), this condition implies that oscillons exist in the theory for the case of real scalar field. 
We conclude that the GSWGC implies existence and formation of oscillons as well as Q-balls.
The left-hand side of Eq.(\ref{eq:sswgcosc}) is the strength of the attractive interaction.  The right-hand side is the parameter that controls the binding energy of two scalars exchanging a scalar mediator. This equation is, therefore, a statement that repulsion is weaker than attraction, which implies that a localized configuration is favored compared with free particle states. 
 
Thus, GSWGC implies that, for a potential with a true vacuum at $\phi = 0$, there exists a Q-ball solution for any value of charge $Q$. 
Small Q-balls with charges $Q$ satisfying 
$\epsilon \equiv Q A^2/ (27 S_\psi m^2) \ll 1$,
where $S_{\psi} \simeq 4.85$~\cite{Linde:1981zj,Brezin:1978}, 
are described by thick-wall approximation
and satisfy $dE/dQ \simeq m (1 - \epsilon^2/2)$~\cite{Kusenko:1997ad}, 
where we assumed Eq.~(\ref{eq:sswgcosc}). 
Large Q-balls, with much larger charge $Q$, can be described by the thin-wall approximation~\cite{Coleman:1985ki}, satisfying
$dE/dQ \simeq m \sqrt{1 - 2 A^2 / (9 \lambda m^2)}$. 
The GSWGC condition \eqref{eq:sswgcosc} implies $dE/dQ \lesssim m \sqrt{1-c'/(6c)}$ in the thin-wall regime. 
When the energy per unit charge, $dE/dQ$, is smaller than the masses of other particles that interact with $\phi$, the Q-ball is absolutely stable and can be a dark matter candidate. If the Q-balls are unstable, as we will discuss below, they can naturally lead to dark matter in the form of PBHs.

An interesting and counter-intuitive example is  the massive scalar field with a running mass, $V(\phi) = \tfrac{1}{2} m^2 \phi^2 (1 + k \ {\rm log} (\phi/M_{\rm pl}))$, where $k$ is a small constant. This potential is approximately realized for flat directions in gravity-mediated supersymmetry-breaking models in the context of MSSM~(e.g.~\cite{Enqvist:1997si, Kasuya:2000sc}). If $k$ is positive, one can check that the GSWGC is satisfied but a Q-ball solution does not exist. However, the potential does not admit a stable vacuum either, because the logarithmic function diverges in the  limit $\phi \to 0$.  For values $\phi \lesssim m$, the potential must be modified so that it has a global minimum at the origin. Then, near the origin, $V(\phi)$ can be expanded as in Eq.~(\ref{eq:mainpot}), in which case the GSWGC implies the existence of a Q-ball solution as we discussed above. 

We have shown that GSWGC suggests existence of solitonic field lumps in the theory. Furthermore, at least for some regions of the parameter space, the GSWGC implies that field instabilities can overcome expansion, allowing for solitons to form in the early universe. 
This has important consequences for cosmology, especially for dark matter.
If the Q-balls are stable (due to a global U(1) symmetry), they constitute a dark matter candidate~\cite{Kusenko:1997si}. 
On the other hand, unstable Q-balls and oscillons   can be the building blocks of primordial black holes, which can be the dark  matter~\cite{Cotner:2016cvr,Cotner:2017tir,Cotner:2018vug,Cotner:2019ykd}. We note that for PBHs formed from scalar field fragmentation the necessary perturbations are decoupled from perturbations generated during inflation. Hence, they can evade potential conflict with swampland conjectures that restrict inflaton potential and inflationary perturbations~\cite{Kawasaki:2018daf}.

In addition to dark matter, our results have implications for Affleck--Dine baryogenesis~\cite{Affleck:1984fy}, where Q-balls play an important role~(see Ref.~\cite{Dine:2003ax} for review). Furthermore, formation of Q-balls or oscillons in the early Universe could generate gravitational waves observable in upcoming experiments~\cite{Kusenko:2008zm,Kusenko:2009cv,Antusch:2016con}.

The work of A.K. and V.T. was supported by the U.S. Department of Energy Grant No. DE-SC0009937.
The work of M. Yamada was supported by an Allen Cormack Fellowship at Tufts University. 
The work of M. Yamazaki was supported by JSPS KAKENHI Grant No. 17KK0087, No. 19K03820 and No. 19H00689.
A.K. and M. Yamazaki were also supported by the World Premier International Research Center Initiative (WPI), MEXT Japan.

\bibliographystyle{apsrev4-1}
\bibliography{scwgcsol}

\end{document}